\documentclass[namedreferences]{solarphysics}

\usepackage[hyperref,optionalrh]{spr-sola-addons} 
\usepackage[optionalrh]{spr-sola-addons} 
\usepackage{epsfig}          
\usepackage{graphicx}        
\usepackage{amssymb}        
\usepackage{color}           
\usepackage{breakurl}                         

\renewcommand{\no}{\noindent}

\newcommand{\aap}{    {\it Astron. Astrophys.}}

\chardef\us=`\_

\begin{document}

\begin{article}
\begin{opening}

\title{Tidal Forcing on the Sun and the 11-year Solar Activity Cycle}

\author[addressref={aff1,aff2},corref,email={gcionco@frsn.utn.edu.ar}]{\inits{R.G.}\fnm{Rodolfo G.}~\lnm{Cionco}\orcid{0000-0002-9032-5815}}
\author[addressref=aff3]{\inits{S.K.}\fnm{Sergey M.}~\lnm{Kudryavtsev}}
\author[addressref=aff4]{\inits{W.-H.}\fnm{Willie}~\lnm{Soon}}
\address[id=aff1]{Comisión de Investigaciones Científicas (CIC), Provincia de Buenos Aires, Argentina}
\address[id=aff2]{Grupo de Estudios Ambientales de la UTN, Colón 332, San Nicolás, Buenos Aires, Argentina}
\address[id=aff3]{M.V. Lomonosov Moscow State University, Sternberg Astronomical Institute, 
13, Universitetsky Pr., Moscow, Russia 
}
\address[id=aff4]{Center for Environmental Research and Earth Sciences (CERES), Salem, MA 01970, USA and 
Institute of Earth Physics and Space Science (ELKH EPSS), H-9400, Sopron, Hungary 
}
\runningauthor{R.G. Cionco et al.}
\runningtitle{Tidal Forcing on the Sun}

\begin{abstract}
The hypothesis that tidal forces on the Sun are related to the modulations
 of the solar-activity cycle has gained increasing attention.
The works proposing physical mechanisms of planetary action 
via tidal forcing have in common that  
quasi-alignments between Venus, Earth, and Jupiter (V-E-J configurations) would provide a 
basic periodicity of $\approx 11.0$ years able to synchronize the operation of
 solar dynamo with these planetary configurations. 
Nevertheless, the evidence behind this particular tidal forcing is still controversial. 
In this context we develop, for the first time, the complete Sun's tide-generating 
potential (STGP) in terms of a harmonic series, where 
the effects of different planets on the STGP are clearly separated and identified.
We use a modification of the spectral analysis method devised by 
\citeauthor{Kudryavtsev2004} ({\it J. Geodesy.} \textbf{77}, 829, \citeyear{Kudryavtsev2004}; 
\aap~\textbf{471}, 1069, \citeyear{Kudryavtsev2007b}) that
 permits to expand any function of planetary coordinates to a harmonic series over 
long time intervals. 
We build a catalog of 713 harmonic terms able to represent the STGP with a high degree of precision.
We look for tidal forcings related to V-E-J configurations and 
specifically the existence of periodicities around $11.0$  years.
 Although the obtained tidal periods range from $\approx$ 1000 years to 1 week, we do not find 
any $\approx$ 11.0 years period. 
The V-E-J configurations do not produce any significant tidal term at this or other periods.
The Venus tidal interaction is absent in the   
 11-year spectral band, which is  dominated by 
Jupiter's orbital motion. 
The planet that contributes the most to the STGP in three planets configurations, 
along with Venus and Earth, is Saturn.
An $\approx 11.0$ years tidal period with a direct physical relevance 
on the 11-year-like solar-activity cycle is highly improbable.
\end{abstract}
\keywords{Solar Cycle, Models; Oscillations; Ephemerides}
\end{opening}

\section{Introduction}
\label{Sec:1}

The hypothesis that a possible tidal forcing on the Sun 
is explicitly related to the modulations 
of the solar-activity cycle has gained increasing attention in the solar--geophysical science community  
\citep[e.g.][]{Scafetta2012, Scafetta2023, Stefani2016, Stefani2019, Stefani2021, Courtillot2021, Charbonneau2022, Nataf2022, Nataf2023, 2022arxiv}. 
Specifically, the works proposing physical mechanisms of 
the planets co-regulating the Sun's magnetic activity
via tidal forcing have in common that 
V-E-J configurations would provide a 
 fundamental periodicity of $\approx 11.0$ years able to synchronize solar dynamo functioning 
with these 
planetary configurations. 
Particularly Stefani and co-authors (cited works) have shown that solar helicity 
oscillations ($\alpha$-mechanism) may be excited with a periodic forcing
 of 11.07 years, 
like the one focused here. Although the physics and origins of the solar cycle are not 
entirely clear, there are advanced enough models (especially for $\alpha$-$\Omega$ dynamos, i.e. 
a dynamo that works with helicity and large-scale differential rotation of
 the magnetized fluids of the Sun) that handle instabilities in the tachocline 
 connected to external parametric forcings.  

The evidence behind the usage of V-E-J configurations as a stable tidal forcing in this
 problem includes 
several estimates and concepts: 
 \rm{a}) the original calculations by 
\cite{Wood1972}, with an obtained period of 11.08  years; 
 \rm{b}) an idea based on ``planetary resonances'' 
\citep[e.g.][]{Scafetta2012, Stefani2021} with a proposed period of 11.07 years;
 \rm{c}) the empirical determinations of  V-E-J quasi-alignments 
   \citep{Okhlopkov2013} with periodicities ranging from 
3.2 years to a main periodicity of $\approx$ 22 years, which implies a ``half spring tidal period''
 of $\approx$ 11 years, etc. 
However, classical spectral analysis applied to approximations of the tide-generating potential on the Sun 
did not find any periodicity related to the $\approx 11.0$ years period   
\citep[][]{Okal&Anderson1975,Nataf2022}.

Venus, Earth, and Jupiter are supposed to be conspicuous or important tidal producers 
but this is not a sufficient condition 
for raising a ``combined'' stable periodic forcing on the Sun.
Dynamical astronomy has a long tradition of considering harmonic perturbations
 with arguments based on combinations of planetary orbital frequencies, $n_i$, of the form 
$k_1 n_1 + k_2 n_2  + \dots$, where $k_i $ are integer multipliers.  
Are these combinations of planetary orbital frequencies always physically significant? 
The answer rests on the development of a 
forcing function that depends on physical and orbital planetary properties. 
For example, 
when analyzing the Sun's barycentric motion using the VSOP87 theory \citep{Bretagnon1988}, 
 the expansion by \citet{Kudryavtsev2009} or the 
EPM2017H ephemeris \citep{CioncoPavlov2018}, 
it is interesting to note that there are several periodicities around 11.07  years; 
for example, $2\,n_5-3\,n_6+2\,n_7$ of  11.042 years period; 
 $n_5+\,n_8$, 11.065  years; $3\,n_5-2\,n_6-8\,n_7$, 11.136  years; 
$2\,n_5-3\,n_6+3\,n_7-2\,n_8$, 11.704  years; etc., 
but all of these arguments are driven by giant planets (sub-indexes 5 -- 8 are assigned 
to planets from Jupiter to Neptune), they do not involve Venus or Earth.  
Although these orbital solutions were not obtained to describe tidal effects,
such earlier results have already hinted that a term associated directly with V-E-J configurations seems to be 
not significant for the Sun's dynamics. 

Taking into account all the reasoning put forward in the preceding paragraphs, we conclude 
that the involvement of V-E-J configurations as a tidal forcing on the Sun is still uncertain.
In order to help to resolve this controversy, we propose to develop the 
STGP in terms of harmonic series.  
Indeed, at present there are no standard harmonic expansions of the STGP where 
the terms caused by the gravitational attraction of different planets (or combinations of them) are clearly separated and identified and which 
 allow one to precisely estimate the absolute values 
of tidal forces acting on elements of the Sun and the sources of these gravitational forces.
 Therefore, there is a real need for an accurate development of the STGP 
 basing on advanced expansion techniques and modern planetary ephemerides, 
 openly available to the scientific community.

\section{Development of the STGP in Terms of Harmonic Series}
\label{Sec:2}

The expression for the STGP [$V(t)$] at an arbitrary point $M$ on the Sun at an epoch $t$ 
is similar to that used in the developments of the tide-generating potential of Earth and other 
terrestrial planets  \citep{Kudryavtsev2004, Kudryavtsev2007a, Kudryavtsev2008}

\begin{eqnarray}
  V(t)=\sum_{j=1}^{8} \mu_j \sum_{n=1}^{n_{\rm{max}}} \frac{r^n}{r_j^{n+1}(t)}P_n\left(\cos \psi_j(t)\right),
	\label{eq:2.1.1}
\end{eqnarray}

\begin{figure}
\centering
	\includegraphics[width=\hsize,height=7.3cm]{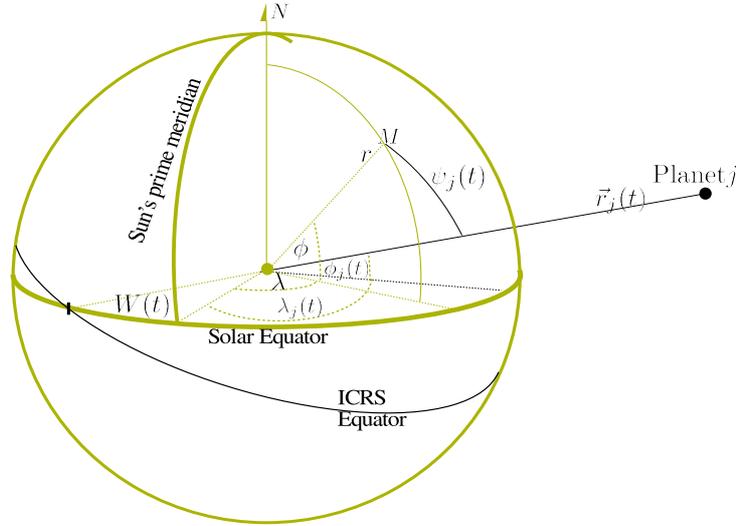}
   \caption{Coordinate system where the STGP was developed (described in  Section~\ref{Sec:2}).
   The Sun's North Pole is marked with $N$.}
    \label{Fig.1}
\end{figure}

\no where $r$ is the heliocentric distance of $M$; $\mu_j=G\,m_j$ is the gravitational parameter 
of the $j^{th}$ attracting body of mass $m_j$, $G$ is the gravitational constant; 
$r_j(t)$ is the heliocentric distance of planet~$j$; $\psi_j(t)$ is an angle between $M$ and planet~$j$ as seen from 
the Sun's center (Figure~\ref{Fig.1}); $P_n$ are the Legendre polynomials of 
degree $n$. 
In our formulation, we considered the effect of all eight major planets on the STGP, 
so that sub-index $j=1$ stands for Mercury, 2 -- for Venus, 3 -- for the 
Earth--Moon barycenter (EMB), 4 -- for Mars, 5 -- for Jupiter, 6 -- for Saturn, 7 -- for Uranus,
 and 8 -- for Neptune. 
 Additionally we considered the separate effects of positions and masses of the Moon and Earth on the STGP, but the difference between the total effect originated from these two bodies and that from the EMB (as a single ``body'') was proven to be negligible within the accuracy of the final series.
Variable $n_{{\rm{max}}}$ defines the maximum degree of the development (to be determined
 experimentally during the expansion procedure).

By using the Legendre addition theorem one can represent  Equation~\ref{eq:2.1.1}
 in the following form

\begin{eqnarray}
V(t) & = & \sum_{n=1}^{n_{{\rm{max}}}} \left(\frac{r}{\rm{R_{\odot}}}\right)^n \sum_{m=0}^n {\bar P}_{nm}(\sin \phi) \nonumber \\
 & \times & \sum_{j=1}^{8} \Phi_{{nm}_j}\left(\lambda,\mu_j,  r_j(t), \phi_j(t), \lambda_j(t), W(t)\right),
\label{eq:2.1.2}
\end{eqnarray}

\no where $\phi$ and $\lambda$ are the heliographic latitude and longitude of $M$, respectively; 
$\rm{R_{\odot}}$ is the Sun's equatorial radius equal to 696\,000 km; 
${\bar P}_{nm}$ are the normalized associated Legendre functions of degree $n$ and order $m$ 
related to the non-normalized ones [$P_{nm}$] as  

\begin{eqnarray}
  {\bar P}_{nm}=\sqrt{\frac{\delta_m (2n+1)(n-m)!}{(n+m)!}} \, P_{nm};   \,
  \delta_m=\left\{
    \begin{array}{l}
     1, \quad {\rm if} \: m=0 \\
     2, \quad {\rm if} \: m \ne 0,  
   \end{array}
   \right.   
	\label{eq:2.1.3}
\end{eqnarray}

\no and $\Phi_{{nm}_j}$ is a function which includes all data related to the 
$j^{\rm{th}}$ attracting planet: $\mu_j$, $r_j(t)$, the heliographic latitude $\phi_j(t)$ and 
the heliographic longitude $\lambda_j(t)$; 
$W(t)$  is the Sun's axial rotation angle reckoned along the 
equatorial plane of the Sun from the ascending node of this
plane on the Equator of the International 
Celestial Reference System (ICRS) to the Sun's prime meridian 
\citep{Archinal2018}. 
The explicit view of $\Phi_{{nm}_j}$-functions can be found in \cite{Kudryavtsev2004, Kudryavtsev2007a}. 

Let us denote
\begin{eqnarray}
\Phi_{nm}(t) \equiv  \sum_j^{8} \Phi_{{nm}_j}\left(\lambda,\mu_j,  r_j(t), \phi_j(t), \lambda_j(t), W(t)\right).
	\label{eq:2.1.5}
\end{eqnarray}

\no Then the task of representing the STGP by harmonic series amounts to 
the development of $\Phi_{nm}(t)$-functions into such series, e.g. with help of a spectral-analysis method. 
In order to initiate the development we first calculated and tabulated
 the values of $\Phi_{nm}(t)$-functions over a long time interval.
As the source of planetary 
heliocentric coordinates we used the latest JPL's long-term 
numerical ephemeris DE-441 \citep{Park2021}. 
The length of the development interval was about 30,000 years (the maximum time interval covered by 
the DE-441 ephemeris)  
and a sampling step of one day was chosen. 
We employ such a long time interval in order to have a possibility to 
look for and identify the expansion terms of large periods in the STGP and 
better separate terms of close-by frequencies.
In particular, the use of a 30,000 years time interval should allow 
us to unambiguously separate and identify several terms with periods close to $\approx$ 11 years 
like those mentioned in Section~\ref{Sec:1}.

At the second step, these tabulated values were processed using a
modification of the spectral analysis method by \cite{Kudryavtsev2004, Kudryavtsev2007b}. 
A key feature of the modified 
method is that it permits the development of a tabulated function to a harmonic series 
where both amplitudes and frequencies of 
the series' terms are not constants but high-degree polynomials of time.  
This feature is important when the development of the STGP over a few thousand years and more is done. 
Over such a long time interval neither planetary motion frequencies nor amplitudes of 
planetary orbital perturbations can be considered as constants, but they are slowly changing 
variables with time \citep[e.g.][]{Bretagnon1988, Simon1994}. 
As a consequence, the development of any function of planetary coordinates 
(including Equations~\ref{eq:2.1.2} and \ref{eq:2.1.5}) to harmonic series 
 carries a similar temporal dependency.  
However, the standard Fourier transform usually indeed assumes constant frequencies and gives constant amplitudes 
of the expansion terms. So that application or adoption of this classical method to development of functions 
of planetary coordinates over thousands and tens of thousands of years 
can lead to significant deterioration of the expansion accuracy.
As a result, over such long time intervals, the harmonic series for the STGP, 
where the amplitudes and arguments of each term are time polynomials, have an explicit advantage. 
They are essentially more compact and accurate than those that could be obtained with use of the classical Fourier transform.

Another feature of the employed method 
is that it directly finds the terms amplitudes at arguments which are linear combinations
 of mean orbital longitudes of the major planets. 
 This allows us to explicitly identify the source of any significant peak in the STGP spectrum.  
More details about the development procedure as well as the accuracy of the method achieved
 for various planetary applications can be found in 
\cite{Kudryavtsev2004, Kudryavtsev2007b,Kudryavtsev2016, Kudryavtsev2017} and \cite{Cionco2021}.

Finally, we obtain the harmonic series for $\Phi_{nm}(t)$ of the form

\begin{eqnarray}
\Phi_{nm}(t)=\sum_{i=1}^{i_{{\rm{max}}}(n,m)} \left[C_{{nm}_i}(t) \cos A_{{nm}_i}(t)+S_{{nm}_i}(t) \sin A_{{nm}_i}(t)\right], 
	\label{eq:2.1.6}
\end{eqnarray}

\no with

\begin{eqnarray}
 C_{{nm}_i}(t) &=& {\rm C0}_{{nm}_i} + {\rm C1}_{{nm}_i} \, t + {\rm C2}_{{nm}_i} \, t^2 + {\rm C3}_{{nm}_i} \, t^3, \nonumber \\
 S_{{nm}_i}(t) &=& {\rm S0}_{{nm}_i} + {\rm S1}_{{nm}_i} \, t + {\rm S2}_{{nm}_i} \, t^2 + {\rm S3}_{{nm}_i} \, t^3,
	\label{eq:2.1.7}
\end{eqnarray}

\no ${\rm C0}_{{nm}_i},\dots,{\rm C3}_{{nm}_i}$, and ${\rm S0}_{{nm}_i},\dots,{\rm S3}_{{nm}_i}$ 
are some derived constants; the time $t$ is reckoned in Julian centuries from the epoch 
J2000.0, that is, at the date of Julian day JD, $t = (\rm{JD}-2451545.0)/36525$; 
the arguments $A_{{nm}_i}(t)$ are defined as 

\begin{eqnarray}
A_{{nm}_i}(t) = m \left[W(t) + \lambda\right] + \sum_{j=1}^{8} k_{ij} l_j (t),
	\label{eq:2.1.8}
\end{eqnarray}

\no where $k_{ij}$ are some obtained sets of integer multipliers; 
 $l_j(t)$ is a temporal polynomial expression for the mean orbital longitude of
 planet $j$ $(j=1,2,\dots,8)$ as given by \cite{Simon1994}. In particular, 
the mean orbital longitudes of Jupiter and Saturn are represented by time polynomials 
of the sixth degree, and similar variables of other six planets are third-degree polynomials
 of time.

When analyzing the effect of a single planet or a linear combination of orbital longitudes of several of 
them on the STGP the following maximum ranges of integer 
multipliers $k_{ij}$ in  Equation~\ref{eq:2.1.8} were used:

\begin{itemize}
\item from -20 to +20 when we estimated the effect of one or a combination of two planets on the STGP; 
\item from -10 to +10 when the effect of three planets was evaluated.
\end{itemize}

The analysis of the final series representing our development of the STGP
reveals that it is sufficient to restrict the maximum number of involved planets to three.
In total we checked for about 200,000 reasonable combinations of integer multipliers $k_{ij}$.
Among them the effects of around 4000 V-E-J configurations on the STGP were analyzed. 
Then for every argument given by  Equation~\ref{eq:2.1.8} 
the amplitudes $C_{{nm}_i}(t)$, $S_{{nm}_i}(t)$ in form of  Equation~\ref{eq:2.1.7} by 
the modified spectral analysis method 
\citep{Kudryavtsev2004, Kudryavtsev2007b} were determined.
 
Let us note that the tidal forces acting on a solar element $M (r, \, \phi, \, \lambda)$ can be straightforwardly 
obtained by using the same set of amplitudes $C_{{nm}_i}(t)$, $S_{{nm}_i}(t)$ and arguments $A_{{nm}_i}(t)$.  
The radial [${F}_r$], latitudinal [${F}_{\phi}$], and longitudinal [${F}_{\lambda}$] tidal forces 
(per unit of mass) at point $M$ are

\begin{eqnarray}
{F}_r &=& \frac{\partial V(t)}{\partial r}, \nonumber \\
{F}_{\phi} &=& \frac{1}{r} \, \frac{\partial V(t)}{\partial \phi}, \nonumber \\
{F}_{\lambda} &=& \frac{1}{r \cos\phi} \, \frac{\partial V(t)}{\partial \lambda}, 
	\label{eq:2.1.9}
\end{eqnarray}

\no where distance $r$ is positive in the direction from the Sun's center; latitude 
$\phi$ is reckoned from the solar equatorial plane being positive to the North; 
and longitude $\lambda$ is counted from the solar prime meridian to the East (Figure 
1). By substituting  Equation~\ref{eq:2.1.2} and  Equation~\ref{eq:2.1.6} to the 
Equations~\ref{eq:2.1.9}, one obtains

\begin{eqnarray}
{F}_r &=& \frac{1}{r}\sum_{n=1}^{n_{{\rm{max}}}} n \, \left(\frac{r}{\rm{R_{\odot}}}\right)^n  
\sum_{m=0}^n {\bar P}_{nm}(\sin \phi)  \nonumber \\
 &\times&  
\sum_{i=1}^{i_{{\rm{max}}}(n,m)} \left[C_{{nm}_i}(t) \cos A_{{nm}_i}(t)+S_{{nm}_i}(t) \sin A_{{nm}_i}(t)\right],   \nonumber  \\
{F}_{\phi} &=&  \frac{1}{r}\sum_{n=1}^{n_{{\rm{max}}}} \left(\frac{r}{\rm{R_{\odot}}}\right)^n  
 \sum_{m=0}^n \frac {{\partial \bar P}_{nm}(\sin \phi)}{\partial \phi}  \nonumber \\
 &\times&  
\sum_{i=1}^{i_{{\rm{max}}}(n,m)} \left[C_{{nm}_i}(t) \cos A_{{nm}_i}(t)+S_{{nm}_i}(t) \sin A_{{nm}_i}(t)\right],  \nonumber  \\
{F}_{\lambda} &=&  \frac{1}{r \cos\phi}\sum_{n=1}^{n_{{\rm{max}}}} \left(\frac{r}{\rm{R_{\odot}}}\right)^n 
 \sum_{m=0}^n  m \, {\bar P}_{nm}(\sin \phi)  \nonumber \\
 &\times&  
\sum_{i=1}^{i_{{\rm{max}}}(n,m)} \left[-C_{{nm}_i}(t) \sin A_{{nm}_i}(t)+S_{{nm}_i}(t) \cos A_{{nm}_i}(t)\right]. 
	\label{eq:2.1.10}
\end{eqnarray}

\no  The derivatives of the associated Legendre functions 
can be easily evaluated using recursive formulas \citep[e.g.][]{AbraStegun}.

\section{Results and Discussion}

An accurate development of the STGP in terms of harmonic series in the 
form given by Equations~\ref{eq:2.1.2}\,--\,\ref{eq:2.1.8} is presented, and
 a corresponding tidal catalog is released at {\textsf{sai.msu.ru/neb/ksm/tgp\_sun/STGP.zip}}.  
The calculations show that the magnitude of the STGP given by
 Equations~\ref{eq:2.1.1} and \ref{eq:2.1.2} can reach values on the order of $10^{-1} \, 
{\rm m}^2{\rm s}^{-2}$. 
As a consequence, the truncation threshold for  the amplitudes of the terms to be included
 in the final STGP series was chosen to be 
 as small as $1\times10^{-7} {\rm m}^2{\rm s}^{-2}$. 
Then, all tidal signal above that limit is considered significant and also identifiable, 
that is, attributable to a specific linear combination of planetary orbital frequencies. 
The maximum degree of the development $n_{{\rm{max}}}$ that leads to terms of such
 minimum amplitudes was found to be equal to 4.
 Table \ref{table1new} presents the number of terms obtained for every value 
 of degree $n$ and order $m$ of the development.
In total, the STGP catalog includes 713 harmonic terms. 

\begin{table*} [!ht]
\caption{Number of terms in the STGP development of degree $n$ and order $m$.} 
\label{table1new}      
\begin{tabular}{lrrrrr}        
\hline                
 $n$  & \multicolumn{5}{c}{$m$} \\
\cline{2-6} \\
     & 0 & 1 & 2  & 3 & 4 \\ 
 \hline                                   
  2 & 104 & 155 & 304 &  & \\
  3 & 17 & 30 & 31 & 36 & \\ 
  4 & 4 & 7 & 8 & 7 & 10 \\  
 \hline                                   
\end{tabular}
\end{table*}

 In order to make sure that no significant term of the STGP development is missed 
we made the following tests. For every degree $n$ and order $m$ we calculated the 
``residuals function" defined as the differences between the original tabulated 
values of $\Phi_{nm}(t)$-functions and the representation of these values by the obtained 
STGP series in form of  Equation~\ref{eq:2.1.6}. Then we defined a large set of frequencies 
corresponding to periods ranging from 0 to, e.g., several thousand years 
(the upper limit was a free parameter) with a small period step of $1\times10^{-4}$ year. 
Finally, we made the Fourier transform of the ``residuals function" at every 
frequency from that set and found the amplitude of the corresponding term. 
If the term amplitude exceeded the chosen truncation threshold we tried to identify a 
combination of the planetary orbital frequencies that has the same or very close period 
and add it to our development of the STGP. 
In this way we could eventually make sure  that all significant terms are captured by 
our STGP series. 
 
Table \ref{table2new} gives the number of terms in the STGP development which include
the orbital longitudes of the various number of planets in the terms arguments.
When the number of involved planets is zero it means that the corresponding term 
is purely associated with the solar rotation or it is a constant.

\begin{table*} [!ht]
\caption{The number of planets used in calculation of arguments of the STGP terms.} 
\label{table2new}      
\begin{tabular}{cr}        
\hline                
 Number of planets   & Number of terms \\
 \hline                                   
  0 & 8 \\
  1 & 307 \\ 
  2 & 313 \\ 
  3 & 85 \\
 \hline                                   
\end{tabular}
\end{table*}
 
Table \ref{table3new} shows how many times the orbital longitude of every  planet 
is used in the calculation of arguments of the STGP terms. 

\begin{table*} [!ht]
\caption{Usage of orbital longitudes of planets in arguments of the STGP terms.} 
\label{table3new}      
\begin{tabular}{lr}        
\hline                
 Planet   & Number of times \\
 \hline                                   
  Mercury & 157 \\
  Venus & 115 \\ 
  EMB & 100 \\ 
  Mars & 45 \\ 
  Jupiter & 334 \\ 
  Saturn & 283 \\ 
  Uranus & 134 \\ 
  Neptune & 20 \\ 
 \hline                                   
\end{tabular}
\end{table*}

A selection of the STGP's terms and associated quantities   
is presented in Table \ref{table1}. 
The amplitude [$Q$] and period [$T$] of every term 
are calculated at the J2000.0 epoch:

\begin{eqnarray}
Q=\sqrt{ {\rm C0^2}_{{nm}_i}+ {\rm S0^2}_{{nm}_i} },
	\label{eq:2.1.13}
\end{eqnarray}

\begin{equation}
T = \frac{2\,\pi}{\dot{A}_{{nm}_i}(t)} 
\end{equation}
   
\no where $ \dot{A}_{{nm}_i}(t) = m \, \dot{W}(t) + \sum_{j=1}^{8} k_{ij} \dot{l}_j (t)$.
The values of $T$- and $Q$-parameters at another epoch can be obtained 
from the complete polynomial expressions for $C_{{nm}_i}$-,$S_{{nm}_i}$-coefficients and 
the terms arguments available in the on-line version of the STGP catalog (\textsf{STGP.zip} file).
The exact calculation procedure and all necessary polynomial expressions for the 
terms arguments and other involved variables are given in the {\textsf ReadMe} file included 
in the \textsf{STGP.zip} archive.
When all integer multipliers $k_{ij}$ are equal to zero, it means that the corresponding term 
is either a constant or its period $T$ is 
due to the Sun's rotation only (e.g. terms with ranks 282 and 613).
Table~\ref{table1} includes all the terms with a period between 10 years and 12 years  
which we identify as the 
11-year spectral band; in addition, the most important terms 
(i.e. with the largest $Q$-values) involving Venus, Earth, or Jupiter are reported. 
The terms are given in the decreasing order of their periods.
 
\begin{table*} [!ht]
\caption{Selected terms of the STGP harmonic development ranked by decreasing period. 
Amplitudes [$Q$] and periods [$T$] are given at the epoch J2000.0. 
Periods are given in years or days (if marked by an asterisk as a superscript).} 
\label{table1}      
\begin{tabular}{rrrrrr}        
\hline                
 Rank & $n$  & $m$ & Planets Involved &  $T$ [yr, d\uppercase{*}] & $Q$ [$\times 10^{-10}$ m$^2$s$^{-2}$] \\    
\hline
    4 & 2 & 0 & $ l_5    -7  l_7   $        &   1010.1953 & $0.1516 \times 10^{5}$ \\
    5 & 2 & 0 & $ -2  l_5  + 5  l_6    $    &    883.2639 & $0.2465\times 10^{5}$ \\
    6 & 2 & 0 & $                      l_8$ &    164.7701 &$ 0.2123\times 10^{4}$ \\
    7 & 2 & 0 & $ -l_3  + 12  l_5         $ &     85.8175 & $0.1420\times 10^{4}$ \\
   10 & 2 & 0 & $  l_5  -2  l_6   $        &     60.9469 & $0.6607\times 10^{5}$ \\
   19 & 2 & 0 & $  2  l_6  -2  l_7        $ &     22.6801 & $0.1523\times 10^{4}$ \\
   20 & 2 & 0 &   $  l_5  -l_6            $ &   19.8589 & $0.1446\times 10^{6}$ \\
   21 & 2 & 0 &   $ -l_5  + 4  l_6        $ &   19.4222 & $0.1861\times 10^{4} $\\
   27 & 2 & 0 &   $  2  l_5 -5 l_6 + 7 l_7$ &   12.1683 & $0.6876\times 10^{4} $\\
   28 & 2 & 0 &   $  3  l_5  -5  l_6      $ &    12.0235 & $0.2836\times 10^{6}$ \\
   29 & 2 & 0 &   $  7  l_7   $             &     12.0029 & $0.3874\times10^{5}$ \\
   30 & 2 & 0 &   $ l_1    -4  l_3   -2  l_6 $ & 11.8774 & $0.4073\times10^{4} $\\
   31 & 2 & 0 &   $ l_5                   $ &      11.8620 & $0.4179\times10^{8}$ \\
   32 & 3 & 0 &   $ l_5                   $ &      11.8620 & $0.6946\times10^{5} $\\
   33 & 2 & 0 &   $ 3  l_5  -6  l_6   + 3l_7 $ &  11.7746 & $0.1448\times10^{4} $\\
   34 & 2 & 0 &   $ 2  l_5   -7  l_7      $ &     11.7243 & $0.3271\times10^{5} $\\
   35 & 2 & 0 &   $ -l_5 + 5  l_6         $ &     11.7048 & $0.9212\times10^{5 }$\\
   36 & 2 & 0 &   $ 4  l_5  -7  l_6  $      &     10.0423 & $0.6635\times10^{4 }$\\
   37 & 2 & 0 &   $ 2  l_5  -2  l_6   $     &      9.9294 & $0.4675\times10^{6 }$\\
   40 & 2 & 0 &   $ -l_2 +  2l_3  -8l_6 $ &      9.7192 & $0.1013\times10^{4 }$\\
   49 & 2 & 0 &   $  l_1   -4  l_3        $ &      6.5751 & $0.8065\times10^{5}$ \\
   76 & 2 & 0 &   $ l_2  -l_3             $ &      1.5987 & $0.2203\times10^{4}$ \\
   80 & 2 & 0 &   $  l_3                  $ &      1.0000 & $0.6374\times10^{7}$ \\
   83 & 2 & 0 &   $  2  l_2  -2  l_3      $ &    291.9607\uppercase{*} & $0.2469\times10^{5}$ \\
   84 & 2 & 0 &   $ 4  l_2  -5  l_3       $ &    243.1650\uppercase{*} & $0.3184\times10^{4 }$\\
   85 & 2 & 0 &   $ l_2  -l_5             $ &    236.9919\uppercase{*} & $0.5996\times10^{4 }$\\
   87 & 2 & 0 &   $ l_2                   $ &    224.7008\uppercase{*} & $0.5607\times10^{7 }$\\
   92 & 2 & 0 &   $ 3  l_2  -3  l_3       $ &    194.6405\uppercase{*} & $0.1491\times10^{5 }$\\
   107& 2 & 0 &   $ 2  l_2    -2  l_5     $ &   118.4960\uppercase{*} & $0.2507\times10^{4 }$\\
   282& 2 & 1 &   $    0                  $ &     25.3800\uppercase{*} & $0.1210\times10^{9 }$\\
   300 &2 & 1 &  $   l_5                  $ &     25.2322\uppercase{*} & $0.3882\times10^{7}$ \\
   359& 2 & 2 &  $-2l_1                   $ &     17.8358\uppercase{*} & $0.1903\times10^{9 }$\\
   406& 2 & 2 &   $ -2  l_2               $ &     14.3059\uppercase{*} & $0.4799\times10^{9 }$\\
   430& 2 & 2 &   $ -2  l_3               $ &     13.6376\uppercase{*} & $0.2242\times10^{9 }$\\
   544& 2 & 2 &   $  -2  l_5              $ &     12.7648\uppercase{*} & $0.4967\times10^{9 }$\\
   555 &2 & 2 &   $ l_2  -2  l_3   +  7  l_6 $ &     12.7506\uppercase{*} & $0.1862\times10^{5}$ \\
   613& 2 & 2 &   $0$                       &     12.6900\uppercase{*} & $0.5515\times10^{7}$ \\
 \hline                                   
\end{tabular}
\end{table*}

Although the obtained tidal periods range from $\approx$ 1000 years to 1 week, we do not find 
any $\approx$ 11.0 years period. 
The V-E-J configurations do not produce any significant tidal term at this or other periods.
No term with an $\approx 22.0$ years period is found either. 
The 11-year spectral band is  dominated by 
Jupiter's orbital motion (terms with rank 31 and 32), 
followed by a combined term originated from both Jupiter and Saturn 
 motions (rank 35). 
No term due to Venus is found in the 11-year spectral band: the term arisen 
from the argument $-l_2+2l_3-8l_6$ (rank 40) with a period of 9.7192 years is the closest one.
The planet that contributes the most to the STGP in three-planets configurations, 
along with Venus and Earth, is Saturn (e.g. rank 40, 555).
Mercury is involved in several periods larger than 1 year (i.e. larger than its orbital period), and 
especially in the 11-year spectral band (rank 30).

In general, the more planets constitute the argument of an STGP term, the less should be the term's amplitude and its effect on solar tides. 
To show this, let us note that the main form of the tide-generating potential on the planet's (or Sun's) surface coincides
with the form of the disturbing function acting on a planet's satellite from other attracting planets \citep[see, e.g.,][]
{Musen1961, Kaula1962}, if we assume the height of the satellite above the planetary surface is equal (just formally) to zero.
Analytic representation of the satellite motion reveals arguments which can
simultaneously include orbital frequencies of two or three (or more) planets. 
It happens when we calculate the satellite orbital perturbations of the second or 
third (or higher) order, respectively. 
However, it is well known that amplitudes of higher-order terms are in general much less than those of lower-order terms 
(if there are no resonant arguments). 
Then, it is expected that an STGP term with an argument including orbital frequencies of three planets 
(such us Venus, Earth, and Jupiter) 
should also have an essentially weaker effect on the STGP and solar tides than terms of 
about the same frequency but originating from motions of one or two planets.

\section{Conclusions}
\label{sec:4}
Various V-E-J configurations do not produce 
any significant term in the STGP harmonic development. 
An $\approx 11.0$ years tidal period with a direct physical relevance  
to the 11-year-like solar-activity cycle is highly improbable.
We can conclude that a combined effect of three (and more) planets
should have a much weaker effect on the STGP than the effect of one or two planets.

The solar barycentric movement was already studied by using current methods 
of celestial mechanics, both analytically \citep{Bretagnon1988} and
 numerically \citep{Kudryavtsev2009, CioncoPavlov2018}. 
Now we complete the study of solar barycentric dynamics with the standard development of the 
STGP in terms of harmonic series, offering a general solution for calculating the tidal 
forcing on the Sun. 
We present this research tool to the scientific community interested in these topics and propose 
an a priori evaluation of the tidal effect of the major planets on the 
Sun to avoid confusions about the relevance of various periodic terms or even spurious forcings.

\begin{fundinginformation}
This research received no specific grant from any funding agency.
\end{fundinginformation}

\begin{dataavailability}
The full output (713 terms) of the STGP catalog is available at \\ 
\textsf{sai.msu.ru/neb/ksm/tgp\_sun/STGP.zip}
\end{dataavailability}

\begin{ethics}
\begin{conflict}
The authors declare that they have no conflicts of interest.
\end{conflict}
\end{ethics}

\bibliographystyle{spr-mp-sola.bst}      
\bibliography{references}    %
\end{article} 

\end{document}